\documentstyle[prd,aps,12pt]{revtex}

\title{
Separability of Rotational Effects on a Gravitational Lens
}

\author{
Hideki Asada,$^{}$\footnote{Electronic address:
asada@phys.hirosaki-u.ac.jp} 
Masumi Kasai$^{}$\footnote{Electronic address:
kasai@phys.hirosaki-u.ac.jp} 
and Tatsuya Yamamoto 
}
\address{
Faculty of Science and Technology, Hirosaki University, 
Hirosaki 036-8561, Japan 
} 
\date{\today}

\begin{document}
\maketitle

\begin{abstract} 
We derive the deflection angle up to $O(m^2a)$ due to a Kerr 
gravitational lens with mass $m$ and specific angular momentum $a$. 
It is known that at the linear order in $m$ and $a$ 
the Kerr lens is observationally equivalent to the Schwarzschild one 
because of the invariance under the global translation of the center 
of the lens mass. We show, however, nonlinear couplings break 
the degeneracy so that the rotational effect becomes in principle 
separable for multiple images of a single source. 
Furthermore, it is distinguishable also for each image of 
an extended source and/or a point source in orbital motion. 
In practice, the correction at $O(m^2a)$ becomes $O(10^{-10})$ 
for the supermassive black hole in our galactic center. 
Hence, these nonlinear gravitational lensing effects are 
too small to detect by near-future observations. 
\end{abstract}

\begin{flushleft}
PACS Number(s); 95.30.Sf, 98.62.Sb, 04.70.Bw
\end{flushleft}

\section{Introduction}
It is of great importance to elucidate the nature of compact objects 
like black holes and neutron stars. In particular, the general
relativity predicts the frame-dragging effect around rotating
objects, which has not been detected. A way of studying 
the rotational effect of the curved spacetime is measuring the 
light propagation as well as monitoring satellite motion. 
As for the gravitational lensing caused by rotating objects \cite{ES}, 
it is known that at the linear order the rotational effect is not 
distinguishable from the translation 
of the center of the lens mass \cite{BHN,AK}. 
In other words, the Kerr lens would be equivalent to the Schwarzschild 
lens without any knowledge of the precise position of the lens \cite{AK}. 
Can {\it nonlinear} effects break the degeneracy 
between the Schwarzschild and Kerr lenses ? 
The main purpose of the present paper is to answer this. 
We will assume a considerably compact object to take account of 
a coupling between the angular momentum and the mass. 
Actually, recent observations \cite{DMD,SHM} have suggested, 
there might be in our universe very compact objects like 
a {\it quark} star whose radius is several kilometers, about a half 
of that of a neutron star, though some arguments are still going on 
\cite{WL}. 

The light propagation in the Kerr spacetime was formulated 
by using the constants of the null geodesics in polar
coordinates \cite{CB72,CB73,BPT,Carter}. However, the approach is not 
suitable for description of the gravitational lens, which is a mapping 
between 2-dimensional vectors on lens and source planes \cite{SEF}. 
Hence, we follow another approach developed recently 
for the gravitational lens \cite{AK}. 

\section{Formulation of the Stationary Gravitational Lens}
First, we summarize notations and equations for gravitational
lensing. We basically follow the notation of Ref.~\cite{SEF}, but the 
signature is $(-,+,+,+)$. 
It is convenient to express the metric of a stationary spacetime 
in the following form: 
\begin{eqnarray}
ds^2&=&g_{\mu\nu} dx^{\mu}\,dx^{\nu} \nonumber\\
&=&-h \left( cdt-w_idx^i \right)^2 + h^{-1} 
\gamma_{ij} dx^i\,dx^j, 
\label{eq:llform}
\end{eqnarray}
where 
\begin{equation}
  \label{eq:components}
  h \equiv - g_{00}, \quad w_i \equiv - \frac{g_{0i}}{g_{00}},  
\end{equation}
and 
\begin{equation}
  \label{eq:gammaij}
  \gamma_{ij} dx^i\,dx^j \equiv -g_{00}
  \left(g_{ij} + \frac{g_{0i}\, g_{0j}}{g_{00}}\right)dx^i\,dx^j
  \equiv d\ell^2. 
\end{equation}
This is essentially the same as the Landau-Lifshitz $3+1$
decomposition of a stationary spacetime \cite{LL}. One difference is the
definition of the spatial metric. They use
\begin{equation}
  \label{eq:llgamma}
  \tilde{\gamma}_{ij} \equiv \left(g_{ij} + \frac{g_{0i}\,
  g_{0j}}{g_{00}}\right)
  = h^{-1} \gamma_{ij}
\end{equation}
as the spatial metric.  
We will hereafter use the conformally rescaled $\gamma_{ij}$, 
since the spatial distance $d\ell$ defined by Eq.~(\ref{eq:gammaij}) 
behaves as the affine parameter of the null geodesics in this 
spacetime \cite{AK}. 
The conformal factor $h$ corresponds to the gravitational redshift
factor.


For a future-directed light ray, the null condition $ds^2 = 0$ gives
\begin{equation}
  c\, dt = \frac{1}{h} \sqrt{\gamma_{ij}dx^i dx^j}
        + w_i \, dx^i . 
\end{equation}
Since the spacetime is stationary, $h, \gamma_{ij}$, and $w_i$ are
functions only of the spatial coordinates $x_i$. 
Then, the arrival time of a light ray is given by the integration from the
source to the observer denoted by the subscript $S$ and $O$,
respectively:  
\begin{equation}
  t \equiv \int_{t_S}^{t_{O}} dt = \frac{1}{c} \int_{S}^{O}
  \left(\frac{1}{h}\sqrt{\gamma_{ij}e^i e^j}
        + w_i \,e^i\right) d\ell, 
\end{equation}
where
$e^i = {dx^i}/{d\ell}$ is the unit tangent vector along the light ray. 
Hereafter, lowering and raising the indices of the
spatial vectors are done by $\gamma_{ij}$ and its inverse $\gamma^{ij}$. 

Fermat's principle states $\delta t = 0$, which provides 
the Euler-Lagrange equation for the light ray,  
fully valid in any stationary spacetime 
\begin{equation}
\label{eq:fulldedl}
  \frac{de^i}{d\ell} = - \left(\gamma^{ij} - e^i e^j\right) \partial_j 
  \ln h - \gamma^{il}\left(\gamma_{lj,k} - \frac{1}{2}
  \gamma_{jk,l}\right) e^j e^k + h \gamma^{ij}
  \left(w_{k,j}-w_{j,k}\right) e^k. 
\end{equation}


The deflection angle $\bbox{\alpha}$ is defined as the difference
between the ray directions at the source ($\ell=-\infty$) and 
the observer ($\ell=\infty$) in the asymptotically flat regions, 

\begin{eqnarray}
\bbox{\alpha} &\equiv& \bbox{e}_{S} - \bbox{e}_{O} 
\nonumber\\
&=&-\int^{\infty}_{-\infty} d\ell \frac{d\bbox{e}}{d\ell}. 
\label{eq:defangle}
\end{eqnarray}


The lens equation relates the angular position of the image 
$\bbox{\theta}$ to the source angular position $\bbox{\beta}$ 
\begin{equation}
  \bbox{\beta} = \bbox{\theta} - \frac{D_{LS}}{D_{OS}}
                                     \bbox{\alpha}(D_{OL}\bbox{\theta}), 
\end{equation}
where $D_{OS}$ is the distance from the observer to the source,
$D_{OL}$ is from the observer to the lens, and $D_{LS}$ is from the
lens to the source, respectively. The vectors $\bbox{\alpha}$, 
$\bbox{\beta}$ and $\bbox{\theta}$ are 2-dimensional vectors in the
sense that they are orthogonal to the ray direction $\bbox{e}$ within
our approximation. In a cosmological situation, 
the unlensed position $\bbox{\beta}$ is not an observable, 
because we cannot remove the lens from the observed position. 

We choose the origin of the spatial coordinate as the location 
of the lens. 
We use a freedom in choosing the origin of $\ell$, so that 
the closest point of the light ray to the lens, denoted by 
$\xi^i$, can be set at $\ell=0$, namely $\xi^i=x^i(\ell=0)$. 
We denote the tangential vector at the closest point 
by $\bar e^i \equiv e^i(\ell=0)$. 
The impact parameter $\bbox{b}$ is the distance from the lens 
to a {\it fiducial} straight line $\bbox{\bar x}(\ell)$ 
which is the tangent to the light ray at the observer, 
while the impact parameter is defined usually at the emitter 
in the standard context of the classical mechanics: 
This is due to the geometrical configuration from which 
the lens equation for $\bbox{\theta}=\bbox{b}/D_{OL}$ 
is derived \cite{SEF}. 
Hence, the impact parameter $\bbox{b}$ is defined as 
\begin{equation}
\bbox{b}=\bbox{\bar x}(\ell=0) . 
\label{eq:impact}
\end{equation}

\section{Gravitational lensing in the Kerr spacetime}
For a slowly rotating case, the Kerr metric is written as 
\begin{eqnarray}
ds^2&=&-\Bigl(1-\frac{2m}{r}\Bigr)dt^2
-\frac{4ma\sin^2\theta}{r}dtd\phi
\nonumber\\
&&+\frac{dr^2}{1-\frac{2m}{r}}+r^2(d\theta^2+\sin^2\theta d\phi^2) 
+O(a^2) , 
\label{eq:kerr} 
\end{eqnarray}
where we used the units of $G=c=1$.  

In order to change this metric into a spatially {\it isotropic} form, 
we perform a coordinate transformation as 
\begin{equation}
r=R(1+\frac{m}{2R})^2 , 
\end{equation}
so that we obtain 
\begin{eqnarray}
ds^2&=&-\Bigl(\frac{1-\frac{m}{2R}}{1+\frac{m}{2R}}\Bigr)^2 
\Bigl(dt
+\frac{2m(\bbox{a}\times\bbox{x})\cdot
  d\bbox{x}}{R^3(1-\frac{m}{2R})^2}  
\Bigr)^2
\nonumber\\
&&+\Bigl(1+\frac{m}{2R}\Bigr)^4 d\bbox{x}\cdot d\bbox{x}
+O(a^2) , 
\label{eq:kerr3} 
\end{eqnarray}
where we introduced a 3-dimensional vector notation 
\begin{eqnarray}
&&\bbox{x}=(x, y, z)=(R\sin\theta\cos\phi, R\sin\theta\sin\phi, 
R\cos\theta) , \\
&&\bbox{a}=(0, 0, a) . 
\end{eqnarray} 

The correspondence between the metric and our (3+1) expression 
given by Eq. $(\ref{eq:llform})$ is 
\begin{eqnarray}
&&h=\Bigl(\frac{1-\frac{m}{2R}}{1+\frac{m}{2R}}\Bigr)^2+O(a^2) , \\
&&\bbox{w}=-\frac{2m(\bbox{a}\times\bbox{x})} 
{R^3(1-\frac{m}{2R})^2}+O(a^2) , \\
&&\gamma_{ij}=\Bigl(1-\frac{m^2}{4R^2})^2\delta_{ij}+O(a^2) , 
\end{eqnarray}
where $\delta_{ij}$ is Kronecker's delta. 
It is worthwhile to note 
\begin{equation}
\frac{\partial}{\partial x^j}\ln h = 
\frac{2mx^j}{R^3}+O(a^2, m^3) . 
\end{equation}

A condition for the closest point is expressed as  
\begin{equation}
\left. \frac{d}{d\ell} (\gamma_{ij}x^ix^j) \right|_{\ell=0}=0, 
\end{equation}
which means 
\begin{equation}
\bbox{\xi}\cdot\bbox{\bar e}=O(a^2, m^2) , 
\end{equation}
where we used 
\begin{equation}
\gamma_{ij}=\delta_{ij}+O(a^2, m^2) . 
\end{equation}

\subsection{$O(m^0 a^0)$}
The metric is expanded as 
\begin{eqnarray}
&&h=1+O(a^2, m) , \\
&&w_i=O(a^2, m) , \\
&&\gamma_{ij}=\delta_{ij}+O(a^2, m^2) . 
\end{eqnarray}

At the lowest order, Eq. ($\ref{eq:fulldedl}$) is expanded as 
\begin{equation}
\frac{de^i}{d\ell}=O(a^2, m) , 
\end{equation}
which is integrated immediately as 
\begin{equation}
e^i=\bar e^i+O(a^2, m) . 
\end{equation}
Consequently, we obtain a straight trajectory of the light ray as 
\begin{equation}
x^i=\xi^i+\ell \bar e^i+O(a^2, m) . 
\end{equation}
For a later convenience, we define $\xi=|\bbox{\xi}|$ and 
$R_0=\sqrt{|\xi|^2+\ell^2}$.

\subsection{$O(m^1 a^1)$}
Using the parameterization of the photon trajectory 
at the lowest order, we obtain 
\begin{eqnarray}
\frac{de^i}{d\ell}&=&-\frac{2m \xi^i}{R_0^3} 
\nonumber\\
&&
+m \Bigl( 
\frac{4(\bbox{a}\times\bbox{\bar e})^i}{R_0^3} 
+\frac{6\{(\bbox{a}\times\bbox{\xi})\cdot\bbox{\bar e}\}  
(\xi^i+\ell \bar e^i)
-6\ell(\bbox{a}\times\bbox{\xi})^i
-6\ell^2(\bbox{a}\times\bbox{\bar e})^i}{R_0^5} 
\Bigr)
\nonumber\\
&&
+O(a^2, m^2) , 
\end{eqnarray}
which is integrated as 
\begin{eqnarray}
e^i&=&\bar e^i-2m 
\Bigl[ 
\frac{\ell\xi^i}{\xi^2R_0} 
-(\bbox{a}\times\bbox{\bar e})^i 
\left(\frac{\ell}{\xi^2R_0}+\frac{\ell}{R_0^3} \right) 
-\frac{3\{(\bbox{a}\times\bbox{\xi})\cdot\bbox{\bar e}\}\xi^i}{\xi^4} 
\left(\frac{\ell}{R_0}-\frac{\ell^3}{3R_0^3} \right) 
\nonumber\\
&&
+\{(\bbox{a}\times\bbox{\xi})\cdot\bbox{\bar e}\} \bar e^i 
\left( \frac{1}{R_0^3}-\frac{1}{\xi^3} \right)
-(\bbox{a}\times\bbox{\xi})^i
\left( \frac{1}{R_0^3}-\frac{1}{\xi^3} \right)
\Bigr]
+O(a^2, m^2) , 
\label{eq:e11}
\end{eqnarray}
where we used $e^i(\ell=0)=\bar e^i$. 
By integrating this, we obtain the light ray trajectory as 
\begin{eqnarray}
x^i=&&\xi^i+\ell \bar e^i
\nonumber\\
&&-2m \Bigl[\frac{\xi^i(R_0-\xi)}{\xi^2} 
-(\bbox{a}\times\bbox{\bar e})^i  
\left(\frac{R_0-\xi}{\xi^2}-\frac{1}{R_0}+\frac{1}{\xi} \right) 
-\frac{\{(\bbox{a}\times\bbox{\xi})\cdot\bbox{\bar e}\} \xi^i}{\xi^4} 
\left( R_0-\xi+\frac{\ell^2}{R_0} \right)
\nonumber\\
&&
+\frac{\{(\bbox{a}\times\bbox{\xi})\cdot\bbox{\bar e}\} \bar e^i}{\xi^2} 
\left( \frac{\ell}{R_0}-\frac{\ell}{\xi} \right)
-\frac{(\bbox{a}\times\bbox{\xi})^i}{\xi^2} 
\left( \frac{\ell}{R_0}-\frac{\ell}{\xi} \right)
\Bigr]
+O(a^2, m^2) , 
\label{eq:x11}
\end{eqnarray}
where $x^i(\ell=0)=\xi^i$ was used. 

The deflection angle is evaluated as 
\begin{equation}
\bbox{\alpha}=\frac{4m\bbox{\xi}}{\xi^2}
-\frac{4m}{\xi^4}
\Bigl(2\{(\bbox{a}\times\bbox{\xi})\cdot\bbox{\bar e}\}\bbox{\xi} 
+\xi^2(\bbox{a}\times\bbox{\bar e}) \Bigr)
+O(a^2, m^2) . 
\label{eq:alpha11}
\end{equation}
This angle is found to agree with previous results \cite{ES} 
by noticing an identity 
\begin{equation}
\bbox{a}\times\bbox{\bar e}=\frac{\bbox{a}\cdot\bbox{\xi}}{\xi^2}
(\bbox{\xi}\times\bbox{\bar e})
-\frac{(\bbox{a}\times\bbox{\xi})\cdot\bbox{\bar e}}{\xi^2}\bbox{\xi} . 
\end{equation}

\subsection{$O(m^2 a^1)$}
We substitute Eqs. ($\ref{eq:e11}$) and ($\ref{eq:x11}$) 
into Eq. ($\ref{eq:defangle}$). 
After lengthy but straightforward calculations, 
we obtain the deflection angle at $O(m^2a)$ as 
\begin{eqnarray}
\bbox{\alpha}&=&\frac{4m\bbox{\xi}}{\xi^2}
-\frac{4m}{\xi^4}
\Bigl(2\{(\bbox{a}\times\bbox{\xi})\cdot\bbox{\bar e}\}\bbox{\xi} 
+\xi^2(\bbox{a}\times\bbox{\bar e}) \Bigr) \nonumber\\
&&+4m^2 \Bigl[ 
\left(\frac{15\pi}{16}-2\right)\frac{\bbox{\xi}}{\xi^3} 
-\left(\frac{5\pi}{4}-4\right)\frac{\bbox{a}\times\bbox{\bar e}}
{\xi^3} 
\nonumber\\
&&-\left(\frac{15\pi}{4}-10\right)
\frac{\{(\bbox{a}\times\bbox{\xi})\cdot\bbox{\bar e}\}\bbox{\xi}}
{\xi^5} 
\Bigr]
+O(a^2, m^3) . 
\label{eq:alpha21}
\end{eqnarray}
It should be noted that some of the coefficients take a peculiar 
form like $\pi$ plus a rational number. 

Up to this point, we have used $\bbox{\xi}$ which is 
the vector for the closest point. 
We are in the position to consider the impact parameter, 
which is defined at asymptotic regions by Eq. ($\ref{eq:impact}$). 
Asymptotic expansions of Eq. ($\ref{eq:x11}$) for a large $\ell$ 
give us the tangent to the light ray at the observer as 
\begin{eqnarray}
\bbox{\bar x}=&&\bbox{\xi}
+2m \Bigl[\frac{\bbox{\xi}}{\xi}
-\frac{\{(\bbox{a}\times\bbox{\xi})\cdot\bbox{\bar e}\}
\bbox{\xi}}{\xi^3}
+\frac{\bbox{a}\times\bbox{\xi}
-\{(\bbox{a}\times\bbox{\xi})\cdot\bbox{\bar e}\}
\bbox{\bar e}}{\xi^2}\mbox{sgn}(\ell) 
\Bigr]
\nonumber\\
&&
+\ell \Bigl[
\bbox{\bar e}-2m \left(
\frac{\xi^2\bbox{\xi}
-2\{(\bbox{a}\times\bbox{\xi})\cdot\bbox{\bar e}\}\bbox{\xi}
-\xi^2(\bbox{a}\times\bbox{\bar e})}{\xi^4}\mbox{sgn}(\ell) 
+\frac{\bbox{a}\times\bbox{\xi}
-\{(\bbox{a}\times\bbox{\xi})\cdot\bbox{\bar e}\}
\bbox{\bar e}}{\xi^3}
\right)
\Bigr] 
\nonumber\\
&&
+O(a^2, m^2) , 
\label{eq:x11asympt}
\end{eqnarray}
where we denoted a signature function $\ell/|\ell|$ 
by $\mbox{sgn}(\ell)$. 
Substituting this into Eq. ($\ref{eq:impact}$), we obtain 
\begin{equation}
\bbox{b}=\Bigl[1+2m\left(\frac{1}{\xi}
-\frac{(\bbox{a}\times\bbox{\xi})\cdot\bbox{\bar e}}{\xi^3}
\right)\Bigr]\bbox{\xi}
+O(a^2, m^2) , 
\end{equation}
where we used $\mbox{sgn}(0)=0$ .
Hence, we find 
\begin{eqnarray}
\bbox{\xi}=&&\Bigl[1-2m\left(\frac{1}{b}
-\frac{(\bbox{a}\times\bbox{b})\cdot\bbox{\bar e}}{b^3}
\right)\Bigr]\bbox{b}
+O(a^2, m^2) ,  
\end{eqnarray}
where we defined $b=|\bbox{b}|$. 
We substitute this into Eq. ($\ref{eq:alpha21}$), 
so that we obtain 
\begin{eqnarray}
\bbox{\alpha}&=&\frac{4m\bbox{b}}{b^2}
-\frac{4m}{b^4}
\Bigl(b^2(\bbox{a}\times\bbox{\bar e})
-2\{(\bbox{a}\times\bbox{\bar e})\cdot\bbox{b}\}\bbox{b} \Bigr) 
\nonumber\\
&&+4m^2 \Bigl[ 
\frac{15\pi\bbox{b}}{16b^3} 
-\frac{5\pi(\bbox{a}\times\bbox{\bar e})}{4b^3} 
\nonumber\\
&&+\frac{15\pi\{(\bbox{a}\times\bbox{\bar e})\cdot\bbox{b}\}
\bbox{b}}{4b^5} 
\Bigr]
+O(a^2, m^3) .  
\label{eq:alpha212}
\end{eqnarray}

\section{Discussion and Conclusions} 
At $O(ma)$, we find out an infinitesimal translation as \cite{AK}
\begin{equation}
\bbox{\bar b}=\bbox{b}-\bbox{a}\times\bbox{\bar e} , 
\label{eq:translation}
\end{equation}
so that the deflection angle given by Eq. ($\ref{eq:alpha11}$) 
can be rewritten as 
\begin{equation}
\bbox{\alpha}=\frac{4m\bbox{\bar b}}{\bar b^2}+O(a^2, m^2) . 
\end{equation}
This is a global transformation, under which the lens equation 
is invariant. 
As a result, we could not separate the rotational effect 
without independent knowledge of the location of the lens \cite{AK}. 
Namely, lensing properties caused by a Kerr lens, such as 
the image positions, magnifications and time delay, 
could be reproduced by a Schwarzschild lens at the suitable position. 

At the next order, we can discover an infinitesimal transformation as 
\begin{eqnarray}
\bbox{\bar b}&=&\bbox{b}-\bbox{a}\times\bbox{\bar e} 
\nonumber\\
&&-\frac{5\pi m}{16b} 
\left((\bbox{a}\times\bbox{\bar e})-\frac{\{(\bbox{a}\times\bbox{b})
\cdot\bbox{\bar e}\}\bbox{b}}{b^2} \right) , 
\label{eq:transformation} 
\end{eqnarray}
so that the deflection angle in Eq. ($\ref{eq:alpha212}$) 
is rewritten as 
\begin{equation}
\bbox{\alpha}=\frac{4m\bbox{\bar b}}{\bar b^2} 
+\frac{15\pi m^2\bbox{\bar b}}{4 \bar b^3}+O(a^2, m^3) . 
\end{equation}
However, $\bbox{\theta}-\bbox{\beta}$ is not invariant under 
this {\it local} transformation. 
As a consequence, the lens equation is not invariant, so that we can 
distinguish the rotational effect on {\it multiple} images of 
a {\it point} source, such as changes in relative positions of images.    
Furthermore, we can recognize it for an {\it extended} source 
(e.g. spherical stars, binary stars and luminous discs) 
and even for a point source if it {\it moves} for instance 
on a straight line or a Keplerian orbit. 

In order to illustrate the rotational effects on the relative 
separation between images, let us consider the lens equation 
in the unit of the Einstein ring radius angle as 
\begin{eqnarray}
\bbox{\theta}_S&=&\bbox{\theta}_I-\frac{\bbox{\theta}_I}{\theta_I^2}
-\lambda\frac{\bbox{\theta}_I}{\theta_I^3} \nonumber\\
&&
+\lambda\Bigl(\frac{\bbox{s}\times\bbox{\bar e}}{\theta_I^2} 
-\frac{2\{(\bbox{s}\times\bbox{\bar e})
\cdot\bbox{\theta}_I\}\bbox{\theta}_I}{\theta_I^4} \Bigr) 
\nonumber\\
&&
+\frac43\lambda^2
\Bigl[ 
\frac{\bbox{s}\times\bbox{\bar e}}{\theta_I^3} 
-\frac{3\{(\bbox{s}\times\bbox{\bar e})\cdot\bbox{\theta}_I\}
\bbox{\theta}_I}{\theta_I^5} 
\Bigr]
\nonumber\\
&&
+O(s^2, \lambda^3) , 
\label{m2a-lenseq}
\end{eqnarray}
where we defined \cite{AK} 
\begin{eqnarray}
\theta_E&=&\sqrt{\frac{4mD_{LS}}{D_L D_S}} , \\
\bbox{\theta}_S&=&\frac{\bbox{\beta}}{\theta_E} , \\
\bbox{\theta}_I&=&\frac{\bbox{\theta}}{\theta_E} , \\
\lambda&=&\frac{15 \pi D_S \theta_E}{64D_{LS}} , \\
\bbox{s}&=&\frac{16}{15\pi}
\frac{\bbox{a}-(\bbox{a}\cdot\bbox{\bar e})\bbox{\bar e}}{m} , \\ 
s&=&|\bbox{s}| . 
\end{eqnarray}
Here, it should be noted that the rotational effect 
comes from $\bbox{s}$ which is  proportional to the projection 
of the spin vector onto the lens plane. 
For a nearby stellar mass black hole and a supermassive one 
in our galactic center, the dimensionless parameter $\lambda$ 
becomes respectively 
\begin{eqnarray}
\lambda&\sim&10^{-7}\left(\frac{m}{10M_\odot}\right)^{1/2}
\left(\frac{100\mbox{pc}}{D_S}\right)^{1/2} , \\
\lambda&\sim&10^{-5}\left(\frac{m}{10^6M_\odot}\right)^{1/2}
\left(\frac{8\mbox{kpc}}{D_S}\right)^{1/2} , 
\end{eqnarray}
where we assumed $D_L \sim D_{LS}$. 

For simplicity, we solve perturbatively Eq. $(\ref{m2a-lenseq})$ 
for sources on the equatorial plane, namely 
$\bbox{\beta}\cdot\bbox{a} = 0$.
The solutions which are on the equatorial plane take a form as 
\begin{equation}
\theta_{\pm}=\phi_{\pm}+\lambda\chi_{\pm}+\lambda^2\psi_{\pm}
+O(\lambda^3) , 
\end{equation}
where we defined 
\begin{eqnarray}
\phi_{\pm}&=&\frac12(\theta_S\pm\sqrt{4+\theta_S^2}) , \\
\chi_{\pm}&=&
\frac{(1\pm s)}{\phi_{\pm}\sqrt{4+\theta_S^2}} , \\
\psi_{\pm}&=&\Bigl[
\frac12\theta_S\mp\frac{6+6\theta_S^2+\theta_S^4}
{2(4+\theta_S^2)^{3/2}} 
+\frac{s}{3}\left(\frac{14+6\theta_S^2+\theta_S^4}
{(4+\theta_S^2)^{3/2}} \mp\theta_S \right) 
\Bigr] . 
\end{eqnarray}
Hence, the angular separation between the double images, 
which is one of the important observables, becomes 
\begin{eqnarray}
\Delta\theta &\equiv& \theta_{+} - \theta_{-} \nonumber\\
&=&\sqrt{4+\theta_S^2}
+\lambda\left( 1-s \frac{\theta_S}{\sqrt{4+\theta_S^2}} \right) 
-\lambda^2\left(\frac{6+6\theta_S^2+\theta_S^4}{(4+\theta_S^2)^{3/2}} 
+\frac23 s \theta_S \right)+O(s^2, \lambda^3) . 
\end{eqnarray}
The term of $O(\lambda s)$ can be absorbed into the leading term 
as $\sqrt{4+(\theta_S-\lambda s)^2}$, which corresponds to 
the global translation given by Eq. $(\ref{eq:translation}$). 
The correction due to the terms at $O(m^2a)$ is of the order 
of $\lambda^2 s$, which becomes $O(10^{-10})$ for the supermassive 
black hole in our galactic center even if $s$ is of the order 
of unity.  
It must be interesting to study some models in detail. 
For instance, (1) how light curves change due to a Kerr lens?, 
(2) what changes occur in image positions and motions 
when a source is a binary star particularly a binary pulsar?,  
and (3) how images look like when a source is an accretion disk? 

Our result is in marked contrast to rotational effects 
on the polarization: 
The difference in the polarization angle between {\it double} 
images from a fixed point source appears at the exactly same order 
$O(m^2a)$ \cite{ITT,Nouri}. 
In practice, however, these nonlinear gravitational lensing effects 
are too small to detect by near-future observations 
\cite{OWL,SIM,GAIA,JASMINE}.


\section*{Acknowledgment}
This work was supported in part by the Japanese Grant-in-Aid 
for Scientific Research from the Ministry of Education, Science 
and Culture, No. 13740137 (H.A.) and by the Sumitomo Foundation
(H.A. and M.K.).


\begin{thebibliography}{99}
\bibitem{ES} R. Epstein and I. I. Shapiro, Phys. Rev. D {\bf 22}, 
  2947 (1980). 
\bibitem{BHN} C. Baraldo, A. Hosoya, and T. T. Nakamura, 
  Phys. Rev. D {\bf 59}, 083001 (1999). 
\bibitem{AK} H. Asada and M. Kasai, Prog. Theor. Phys. 
{\bf 104}, 95 (2000). 
\bibitem{DMD}J. J. Drake et al., Astrophys. J. {\bf 572}, 996 (2002).  
\bibitem{SHM}P. Slane, D. J. Helfand and S. S. Murray, 
Astrophys. J. {\bf 571}, L45 (2002). 
\bibitem{WL}F. M. Walter and J. Lattimer, submitted to Astrophys. J. 
(astro-ph/0204199). 
\bibitem{CB72}C. T. Cunningham and J. M. Bardeen, 
Astrophys. J. {\bf 173}, L137 (1972). 
\bibitem{CB73}C. T. Cunningham and J. M. Bardeen, 
Astrophys. J. {\bf 183}, 237 (1973). 
\bibitem{BPT}J. M . Bardeen, W. H. Press and S. A. Teukolsky, 
Astrophys. J. {\bf 178}, 347 (1972).  
\bibitem{Carter}B. Carter, Phys. Rev. {\bf 174} 1559 (1968). 
\bibitem{SEF} P. Schneider, J. Ehlers, and E. E. Falco, {\it
    Gravitational Lenses\/} (Springer-Verlag, Berlin, 1992). 
\bibitem{LL} L. D. Landau and E. M. Lifshitz, {\it The Classical Theory 
    of Fields} (Oxford: Pergamon 1962). 
\bibitem{ITT} H. Ishihara, M. Takahashi, and A. Tomimatsu, 
  Phys. Rev. D {\bf 38}, 472 (1988).
\bibitem{Nouri} M. Nouri-Zonoz, Phys. Rev. D {\bf 60}, 024013 (1999). 
\bibitem{OWL}Overwhelmingly Large Telescope (OWL), 
http://www.eso.org/projects/owl/
\bibitem{SIM}Space Interferometry Mission (SIM), 
http://sim.jpl.nasa.gov/
\bibitem{GAIA}Global Astrometric Interferometer for Astrophysics (GAIA), 
http://astro.estec.esa.nl/GAIA/
\bibitem{JASMINE}Japan Astrometry Satellite Mission 
for INfrared Exploration (JASMINE), http://www.jasmine-galaxy.org/
\end{thebibliography}
\end{document}